\documentclass[twocolumn,preprintnumbers,amsmath,amssymb]{revtex4}
\usepackage{graphicx}
\usepackage{dcolumn}
\usepackage{bm}
\usepackage{braket}

\usepackage{color}
\usepackage{ulem}

\usepackage{here}

\usepackage[version=3]{mhchem}

\font\scripti=cmmi7
\font\scriptscripti=cmmi5
\def\sib#1{\setbox0 = \hbox{\scripti #1}
  \kern-.02em\copy0\kern-\wd0
  \kern.04em\box0} 
\def\ssib#1{\setbox0 = \hbox{\scriptscripti #1}
  \kern-.02em\copy0\kern-\wd0
  \kern.04em\box0} 
\font\tenib=cmmib10 
\skewchar\tenib='177 \skewchar\tenib='177 \skewchar\tenib='177
\def\pbold#1{\setbox0 = \hbox{$ #1 $}
  \kern-.022em\copy0\kern-\wd0
  \kern.011em\copy0\kern-\wd0
  \kern.011em\copy0\kern-\wd0
  \kern.011em\copy0\kern-\wd0
  \kern.011em\box0} 

\usepackage{graphicx}
\usepackage{dcolumn}
\usepackage{bm}

\def\up{\uparrow}

\def\lesssim{\ \raise.3ex\hbox{$<$}\kern-0.8em\lower.7ex\hbox{$\sim$}\ }
\def\gesim{\ \raise.3ex\hbox{$>$}\kern-0.8em\lower.7ex\hbox{$\sim$}\ }

\begin{document}
\title{Resonant pair-exchange scattering and BCS-BEC crossover in a
system\\ composed of dispersive and heavy incipient bands: a Feshbach analogy}
\author{Kazunari Ochi,\(^{1}\) Hiroyuki Tajima,\(^{1, 2}\) Kei Iida,\(^{1}\) and Hideo Aoki\(^{2, 3}\)}
\affiliation{\(^{\textit{1}}\)Department of Mathematics and  Physics, Kochi University, Kochi 780-8520, Japan\\
\(^{\textit{2}}\)Department of Physics, University of Tokyo, Hongo, Tokyo 113-0033, Japan\\
\(^{\textit{3}}\)National Institute of Advanced Industrial Science and Technology (AIST), Tsukuba 305-8568, Japan}
\date{\today}
\begin{abstract}
We theoretically show that a two-band system with very
different masses harbors a resonant pair scattering that leads to novel
pairing properties, as highlighted by the Bardeen-Cooper-Schrieffer (BCS) to
Bose-Einstein condensation (BEC) crossover. Most importantly, the interband
pair-exchange coupling induces an effective intraband attraction in each band,
enhancing the superfluidity/superconductivity.
The effect, a kind of Suhl-Kondo mechanism, is specifically enhanced
when the second band has a heavy mass and is {\it incipient} (lying
close to, but just above, the chemical potential, $\mu$),
which we call a resonant pair scattering.
By elucidating the dependence of the effective interactions and gap functions
on $\mu$, we can draw an analogy between the resonant pair scattering and the Feshbach resonance.
\end{abstract}
\maketitle
\par
\section{Introduction}
One of the central issues in
superconductivity/superfluidity
is the crossover between the Bardeen-Cooper-Schrieffer (BCS)
and Bose-Einstein condensation (BEC) regimes,
or a crossover between weak and strong-coupling
regimes~\cite{Zwerger:2012,Randeria,Strinati,Ohashi}.
Another crucial interest in
recent years is the
multi-band superconductors and superfluids,
which harbor a lot of specific interests.
Indeed there has been
an upsurge of interests in multi-band and
multi-orbital effects on superconductivity in a wide
variety of strongly-correlated solid-state
systems as exemplified by
the iron pnictides, copper oxides, and heavy-fermion compounds~
\cite{AokiSuperconductivity,KurokiRearization,Yamazaki,KurokiSuperconductivity,Mazziotti,Mazziotti:reso}.
Multi-species cold-atom systems have also been
intensively studied for exploring a variety of phenomena.
Now, an intriguing question we want to elaborate
in the present work is: what if we
combine these two subjects to consider
a BCS-BEC crossover in multi-band superconductors and superfluids.
Indeed, in solid-state systems, the iron-based
superconductor is inherently multi-band, and
some compounds in the material family
are considered to be in
a BCS-BEC crossover regime.
In cold-atom systems, there exists, beside
the magnetic Feshbach resonance, what is called
the ``orbital Feshbach resonance" when the
atomic spieces (such as Yb) have inert electron spins but
multiple orbital states. This can be
utilized to provide with open and closed channels
to realize the unitarity-limit
region in the crossover.  Multi-band systems also
give us greater opportunities
in that there are several degrees
of freedom to be engineered, such as
the mass ratio and band offset
between the bands, relative positions
between the chemical potential and the
respective band edges, where we can play around with
inter-band vs. intra-band interactions
in considering superconductivity/superfluidity.
\par

A specific point of interest in multi-band superconductors is
what is called the ``incipient band" situations.
Namely,
in some of the iron-based superconductors,
the hole band has its edge located close to, but slightly
away from, the chemical potential, which is called ``incipient"~\cite{Qian,Tan,Liu,He:2013,Lee,Miao,Niu,Charnukha}.
While the terminology ``incipient" is often
used in the community of the iron-based FeSe
superconductor for the incipient $s_{\pm}$
pairing involving the hole band below $E_{\rm F}$, the
concept of the incipient situation itself
was originally introduced
in a 2005 paper~\cite{narrowwide}.
Namely, the physics is
that the pair scattering mediated by spin fluctuations occurs between the main band and incipient band~\cite{Chen,Linscheid,BangPairing,Kato},
and this can drastically
enhance superconductivity, especially when the
incipient band is flat as found in Ref.~\cite{narrowwide}.
In such situations, the inter-band pair scattering, on top
of the intra-band ones, crucially determins
the gap symmetry~\cite{Suhl,Kondo} (see also Ref.~\cite{AokiJSNM}
for a review).
\par
Further feature in the iron-based superconductors
is that a compound ${\mathrm{Fe}}_{1+y}{\mathrm{Se}}_{x}{\mathrm{Te}}_{1-x}$ realizes crossover from the weak-coupling BCS regime to the BEC condensation of tightly-bound pairs when the iron content $y$ is varied~\cite{Lubashevsky,Okazaki,KasaharaWatashige,KasaharaYmashita}. With decreasing $y$, the hole pocket becomes shallower, which
makes the ratio, $\Delta/E_{\rm F}$, between the superconducting gap and Fermi energy monotonically increase up to 0.5
\cite{Rinot}, which has been regarded as an indication for
the BCS-BEC crossover.
Another solid-state system that accommodates
the BCS-BEC crossover is a hafnium compound ${\mathrm{Li}}_{x}\mathrm{HfNCl}$
tuned with an electric-double-layer structure~\cite{Nakagawa}.
At a low carrier density (\(x=0.04\)), a pseudogap
reminiscent of strong-pairing fluctuations in the BCS-BEC crossover has been observed, with \(\Delta/E_{\rm F}\)
reaching 0.12 at \(x=0.02\).
From theoretical viewpoints,
it has been proposed that similar resonant phenomena
can occur in nanostructures with complicated geometries~\cite{Bianconi}, or in tight-binding band structures~\cite{Avishai}.
\par
If we turn to cold-atom systems,
on the other hand, the unitarity limit
in the BCS-BEC crossover has been intensively
investigated for usual single-orbital, single-species
ultracold Fermi gases~\cite{Regal,Zwierlein},
where $\Delta/E_{\rm F}\simeq 0.4-0.5$ has been reported
~\cite{Schirotzek,Hoinka,Horikoshi}.
In usual cold-atom systems,
typically $^6$Li and $^{40}$K Fermi atomic gases,
are characterized
by the s-wave scattering length $a$ for the
interatomic interaction, which absorbs the ultraviolet
divergence arising from the singular contact-type interaction. The quantity $a$
can be controlled by an external magnetic field
with the magnetic Feshbach resonance associated with the electron-spin degree of freedom with $S=1/2$~\cite{Chin}.
The BCS-BEC crossover is marked by a change of sign of $a$, which is physically quite natural, since it associates the crossover with the formation of a bound state for a pair.
Now,
a realization of the two-band BCS-BEC crossover has
recently been anticipated in
Ytterbium Fermi
gases~\cite{Zhang,Xu,He:2017,Iskin,Mondal}.
In the case of $^{173}$Yb atom with $S=0$, the system
accommodates the orbital Feshbach resonance, which involves intrachannel and interchannel interactions in a two-channel system having different electron-orbital states, \(^{1}S_{0}\) and \(^{3}P_{0}\), and nuclear-spin states
~\cite{Pagano,Hofer,Zhang,Ohashi}.
Corresponding Hamiltonian is similar to the two-band superconductivity model called Suhl-Kondo~\cite{Suhl,Kondo}.
Moreover, a bound-state formation due to the
two-band nature has been demonstrated in recent experiments \cite{Cappellini}.

The Feshbach resonance can also be invoked for inducing the Kondo effect by manipulating the spin exchange interaction in a two-band system~\cite{Gorshkov,Cheng}.
Since the different orbital states of \(\mathrm{^{173}Yb}\) feel different optical-lattice potentials, this can be used to realize a two-band system having different effective masses.
As we shall show, a kind of BCS-BEC crossover occurs in such a system, but that is driven by interband coupling and hence totally different from the usual single-band BCS-BEC crossover, where the scattering length alone is the controllable parameter.

\begin{figure}[t]
\centering{\includegraphics[width=86mm]{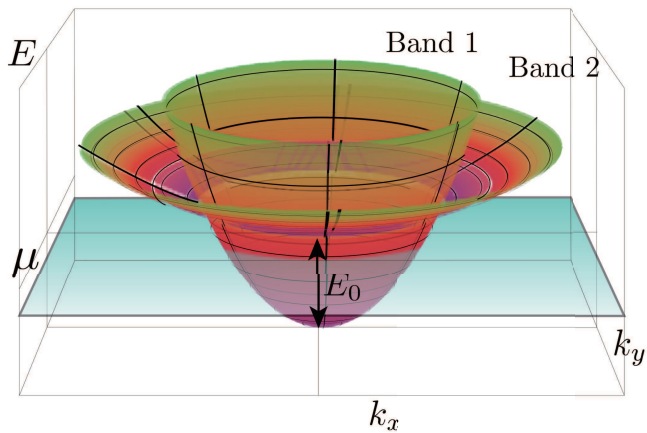}}
\caption{Band dispersions (against $(k_x, k_y)$
with a $k_{z}=0$ projection in this plot) of the two-band system considered here with different effective masses
with a band offset, $E_0$. The upper band (Band 2) is assumed to have a heavier mass than
the lower band (Band 1). Since we are interested in the situation where Band 2 is incipient (see text), the chemical potential $\mu$ is tuned around the bottom of Band 2.}
\label{fig1}
\end{figure}

With the above background, the purpose of the present work is to explore
specific features in the BCS-BEC crossover
that arise when we have a
fermion system (single-species, spin-1/2)
that consists of a lighter-mass band (called
Band 1 hereafter) and a heavier-mass band (Band 2).
We focus on what will happen when
we engineer the system by varying
a band offset, $E_0$, along with the
position of the chemical potential, $\mu$,
on top of the mass ratio of the two bands.
For the reason mentioned above and
elaborated below, we are specifically
interested in the situation when Band 2 is
``incipient", i.e., close to, but detached
from, the chemical potential,
as schematically depcited in Fig.~\ref{fig1}.
The questions we ask ourselves are:
can unusual superconducting or superfluid states
arise when the mass ratio is large in the
presence of intraband and interband pairing
interactions.  The latter gives rise to
interband pair scattering (i.e., virtual
pair-exchange processes across the two bands).
In two-band systems the gap
function has two components,
and we solve the two-component gap equation,
where we focus on the intraband pairing in the case in which the chemical potential is set around the bottom of the incipient band.

We shall particularly clarify how the superfluid/superconducting gaps and number densities behave in the presence of the resonant pair scattering by varying the mass ratio between the two bands.
There, a point of interest is the {\it effective scattering length}~\cite{Ohashi2005,Liu2005,Tajima2018} {\it that characterizes the effective intraband interaction induced by the interband pair scattering}.
We shall show that the superfluid/superconducting gaps in the two bands are strongly enhanced in a manner
drastically dependent on which band.  This
originates from the interband pair scattering when the incipient band is heavy, where the effective scattering lengths
cross from the weak-coupling regime over to the strong-coupling one in a manner drastically dependent on the band.

\par
This paper is organized as follows.
In Sec.~\ref{sec2}, we present the two-band model Hamiltonian and formulate the gap equation to be solved numerically.
We employ the mean-field BCS-Leggett theory
~\cite{Leggett,Zwerger:2012,Ohashi}
, which is known to successfully describe qualitative features of the BCS-BEC crossover at zero temperature in dilute systems as
exemplified by cold atoms.
While the BCS theory basically assumes that the excitation is restricted around the Fermi energy with the density of states taken as a constant, the BCS-Leggett theory employed in this paper includes excitations at shorter wavelengths. Such a difference is crucial for describing the BCS-BEC crossover in the that high-momentum excitations also occur in the strong-coupling regime where the Fermi surface is absent.
This is not directly applicable to the above-mentioned strongly correlated solid-state systems, but is expected to give a hint for the BCS-BEC crossover in multi-band superconductivity.

Within the mean-field theory,
the number density and the effective scattering length in each band are calculated. 
In Sec.~\ref{sec3}, we show numerical results for the chemical potential dependence of the superfluid/superconducting gaps, number density ratio, and effective scattering lengths. 
Section~\ref{sec4} summarizes the paper.
Throughout the text, we use units in which $\hbar=k_{\rm B}=1$, while the system volume is taken to be unity.


\section{Formulation \label{sec2}}
As depicted in Fig.~\ref{fig1}, we consider a two-band continuum model
in three spatial dimensions,
where the bands, with different masses and a band offset,
have dispersions,
\begin{eqnarray}
  \label{eq2}
  \xi_{i}(\bm{k})=\frac{k^{2}}{2m_{i}}-\mu+E_{0}\delta_{i,2}.
\end{eqnarray}
We assume a parabolic dispersion $\xi_{i}(\bm{k})$
against momentum $\bm{k}$
for each band labelled by the index $i=1,2$
having a mass (i.e., an effective mass for
a lattice) $m_{i}$, $E_0$ is the band offset,
and $\mu$ is the chemical potential.
For a given value of $E_0$,
we regard the chemical potential as a control parameter.
The Hamiltonian reads~\cite{Iskin2006}
\begin{eqnarray}
\label{eq1}
H &=& \sum_{i,\bm{k},\sigma}\xi_{i}(\bm{k})c^{\dag}_{\bm{k},\sigma,i}c_{\bm{k},\sigma,i} \cr
  &+&\sum_{i,j}\sum_{\bm{k},\bm{k}^{\prime}}V_{ij}(\bm{k},\bm{k}^{\prime})B^{\dag}_{\bm{k},i}B_{\bm{k}^{\prime},j},\\
B_{\bm{k},j} &=& c_{-\bm{k},\downarrow,j}c_{\bm{k},\uparrow,j},
\end{eqnarray}
where
$c_{\bm{k},\sigma,i}^{\dag}$ creates a fermion with momentum $\bm{k}$ and spin $\sigma (=\up$ or $\downarrow$) in band $i$, and $B^{\dag}_{\bm{k},i}$ is the pair-creation operator in band $i$.
The second term in $H$ describes intraband ($i=j$) and
interband ($i\neq j$) interactions.
For the interaction $V_{ij}$,
we assume in this paper, with cold-atom systems in mind, the contact-type attractive interaction,
\[
V_{ij}(\bm{k},\bm{k}')=-U_{ij}\theta(\Lambda-k)\theta(\Lambda-k'),
\]
where $U_{ij}\geq 0$ and $\Lambda$ is a (spherical) momentum cutoff,
which is required to avoid an ultraviolet divergence due to the contact-type interaction~\cite{Ohashi}.
\par
To renormalize the intraband interaction
$U_{ii}$ against $\Lambda$, we can define, as a measure of $U_{ii}$,  an s-wave intraband scattering length $a_{i}$ in Band $i$ as~\cite{Ohashi},
\begin{eqnarray}
  \label{eq3}
  \frac{4\pi a_{i}}{m_{i}}=\frac{-U_{ii}}{1-U_{ii}\sum^{k\leq \Lambda}_{\bm{k}}\frac{1}{k^{2}/m_{i}+2E_{0}\delta_{i2}}}.
\end{eqnarray}
\par
We apply the mean-field
approximation to both the intra- and inter-band pair scattering processes in the Hamiltonian Eq.(\ref{eq1}) for describing the superfluid/superconducting properties.
The gap equation in a two-band system can be expressed in such a way that the two superfluid/superconducting gaps,
$\Delta_{1}, \Delta_{2}$, are coupled as \cite{Guidini, Yerin}
\begin{eqnarray}
  \label{eq5}
  \Delta_{i}=\sum_{j=1,2}U_{ij}\Delta_{j}\sum_{\bm{k}}^{k\leq\Lambda}\frac{\tanh{\left(\frac{E_{j}(\bm{k})}{2T}\right)}}{2E_{j}(\bm{k})},
\end{eqnarray}
where \(E_{j}(\bm{k}) = [\xi_{j}^{2}(\bm{k})+\Delta_{j}^{2}]^{1/2}\) is the quasiparticle dispersion in the superfluid/superconducting state.
For applying the mean-field approximation, the effect of inter-band pair-scattering processes is non-perturbatively included in our two-band gap equation (\ref{eq5}).
This equation reproduces the two-body bound-state equation in the large interband-coupling limit (see Appendix \ref{ap1}).
In the limit where the interband interactions $U_{12}$ and $U_{21}$ are larger than the intraband interactions $U_{11}, U_{22}$, Eq. (\ref{eq5}) corresponds to the gap equation in Eq. (25) of \cite{Chubukov} where the interband pair scattering is dominant.

We note that Eq.~(\ref{eq5}) can also be obtained from the condition for the gapless collective mode in the \textit{T}-matrix approximation \cite{Ohashi}. Although its form is different from more sophisiticated approaches such as the self-consistent \textit{T}-matrix approximation, we employ the present formalism, since the \textit{T}-matrix approach based on Eq.~(\ref{eq5}) is successfully applied to the BCS-BEC crossover~\cite{L.He, Tajima, Tajima2020}.
\par
Since we are interested in the incipient situation,
we tune $\mu$ around $\mu=E_{0}$ where $\mu$ touches the bottom of Band 2,
in which the occupied number density $n_{i}$ in Band $i$
changes with $\mu$ as
\begin{eqnarray}
  \label{eq6}
  n_{i}=2\sum_{\bm{k}}\left[v^{2}_{i}(\bm{k})f(-E_{i}(\bm{k}))+u^{2}_{i}(\bm{k})f(E_{i}(\bm{k}))\right],
\end{eqnarray}
where \(f(\pm E_{i}(\bm{k}))=1/\left(e^{\pm E_{i}(\bm{k})/T}+1\right)\) is the Fermi-Dirac distribution function, while the BCS coefficients are given as
\begin{eqnarray}
  \label{eq7}
  v^{2}_{i}(\bm{k}) &=& \frac{1}{2}\left[1-\frac{\xi_{i}(\bm{k})}{E_{i}(\bm{k})}\right],\\
  u^{2}_{i}(\bm{k}) &=& 1-v^{2}_{i}(\bm{k}).
  \label{eq72}
\end{eqnarray}
\par

\begin{figure}[t]
\includegraphics[width=80mm]{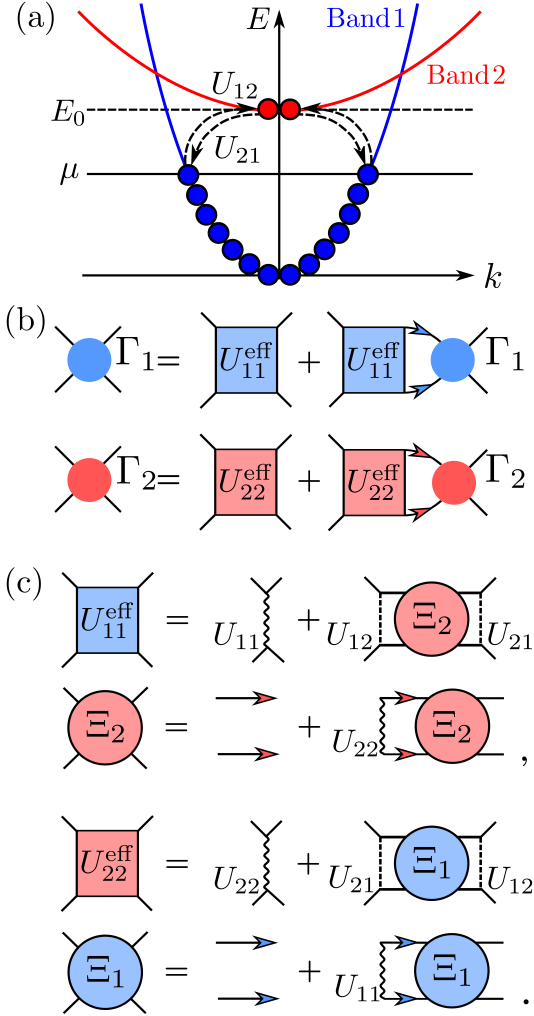}
\caption{(a) Interband pair-scattering processes are
schematically shown on the band dispersion,
here for $\mu<E_{0}$ in the presence of the pair-exchange interaction $U_{12}, U_{21}$. (b) Corresponding diagrams for the interaction vertex $\Gamma_i$ for Band $i$ (Eq.~(\ref{eq8})), which is related with the effective scattering lengths $a_{i}^{\rm eff}$ as defined in  Eq.~(\ref{eq8}). (c) The effective intraband interactions $U_{ii}^{\mathrm{eff}}$ Eq.~(\ref{eq9}),
which are composed of the bare intraband interaction ($U_{ii}$; wavy lines) and the pair-exchange interaction between Band 1 and Band 2, which involves multiple scattering $\Xi_{j}$ in Band $j(\neq i)$.
}
\label{fig2}
\end{figure}

In the presence of the interband interaction $U_{12}$,
we have the resonant pair-scattering, as
shown in Fig.~\ref{fig2} (a)
and captured diagramatically in Fig.~\ref{fig2} (b).
We can then calculate the effective scattering length $a_{i}^{\rm eff}$, which reflects the pair-exchange-induced intraband attraction in Fig.\ \ref{fig2} (b) as
  \begin{eqnarray}
    \label{eq8}
    \frac{4\pi a^{\mathrm{eff}}_{i}}{m_{i}}
\equiv \Gamma_{i}
=\frac{-U^{\mathrm{eff}}_{ii}}{1-U^{\mathrm{eff}}_{ii}\sum^{k\leq \Lambda}_{\bm{k}}\frac{1}{k^{2}/m_{i}+2E_{0}\delta_{i2}}}.
  \end{eqnarray}
Here \(\Gamma_{i}\) is the interaction vertex, and
\(U_{ii}^{\mathrm{eff}}\) is the effective interaction in Band $i$ that can be obtained by rewriting Eq. (\ref{eq5}) as
\begin{eqnarray}
\label{addeq10}
\Delta_{i}=U_{ii}^{\rm eff}\sum_{\bm{k}}^{k\leq\Lambda}\frac{\Delta_{i}}{2E_{i}(\bm{k})}\tanh{\left(\frac{E_{i}(\bm{k})}{2T}\right)}
\end{eqnarray}
with
\begin{eqnarray}
  \label{eq9}
  U_{ii}^{\mathrm{eff}}&=&U_{ii}+U_{ij}\Xi_{j}U_{ji}, \\
  \Xi_{j}&=&\frac{\sum_{\bm{k}}^{k\leq \Lambda}\frac{\tanh
\left(\frac{E_{j}(\bm{k}^{\prime})}{2T}\right)}{2E_{j}(\bm{k})}}{1-U_{jj}\sum_{\bm{k}}^{k\leq \Lambda}\frac{\tanh\left(\frac{E_{j}(\bm{k}^{\prime})}{2T}\right)}{2E_{j}(\bm{k})}}
  \label{eq92}
\end{eqnarray}
for $(i,j) = (1,2)$ or $(2,1)$.
\par
The BCS-BEC crossover is characterized in terms of the dimensionless coupling parameter, $1/(k_{0}a_{i})$, as~\cite{Ohashi}
\begin{eqnarray}
&&1/(k_{0}a_{i})\rightarrow -\infty: {\rm weak\mathchar`-coupling~BCS~limit},\\
&&1/(k_{0}a_{i})\rightarrow +\infty: {\rm strong\mathchar`-coupling~BEC~limit},
\end{eqnarray}
where $k_{0} \equiv \sqrt{2m_{1}E_{0}}$ is the Fermi momentum as defined for a zero-temperature ideal Fermi gas having a mass $m_{1}$ and Fermi energy $E_{0}$.
Since we want to focus on the effects of the pair-exchange coupling,
the intraband couplings are taken to be weak as $1/(k_{0}a_{1})=1/(k_{0}a_{2})=-2$ throughout the present paper.
The crossover of interest here is driven by interband coupling, hence distinct from the usual single-band crossover.
We now examine how the $1/(k_{0}a_{i}^{\rm eff})$ changes
across the BCS to BEC regimes as $\mu$ is increased for
various values of $\tilde{U}_{12}$ and $m_1/m_2$.
The momentum cutoff is here taken to be $\Lambda=100k_{0}$.
We have numerically checked that the result does not
change significantly for larger cutoffs.
\par

\section{Results and Discussions \label{sec3}}
\subsection{Superfluid/superconducting gaps and particle densities}
\begin{figure*}[t]
\centering{\includegraphics[width=132mm]{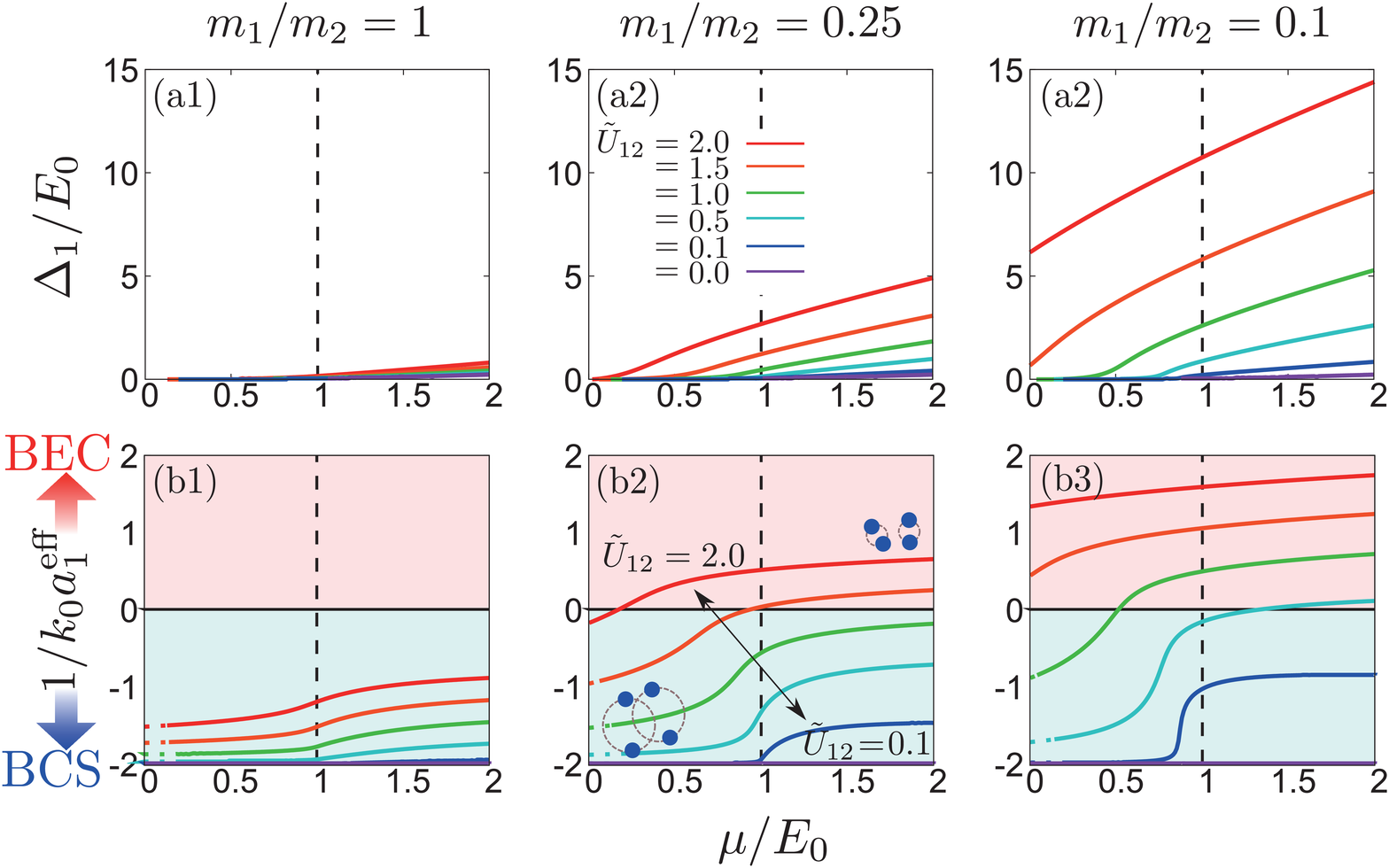}}
\caption{Superfluid/superconoducting gap $\Delta_{1}$ (upper panels) and the inverse effective scattering length \(1/a_{1}^{\mathrm{eff}}\) (lower)
in Band 1 calculated as functions of the chemical potential $\mu$ at different mass ratios
$m_1/m_2=1$ (left panels), $m_1/m_2=0.25$ (middle), and $m_1/m_2=0.1$ (right).  In each panel
the result is obtained for various values of the pair exchange coupling $\tilde{U}_{12}=0.0$, $0.1$, $0.5$, $1.0$, $1.5$, and $2.0$ as color coded.
We take $1/(k_{0}a_{1})=1/(k_{0}a_{2})=-2$.
At $\tilde{U}_{12}=0.0$, $1/(k_{0}a_{1}^{\rm eff})$ coincides with $1/(k_{0}a_{1})=-2$.
The horizontal solid lines at \(1/(k_{0}a_{1}^{\rm{eff}})=0\) represent the unitarity limit, while the vertical dashed lines mark $\mu=E_0$.
For $\mu \rightarrow 0$ where $\Delta_{2}$ is negligibly smaller than $E_{0}$, we display the asymptotic solutions obtained from the two-body calculation at $\mu=0$ (see Appendix~\ref{ap1}) as dotted curves.}
\label{fig3}
\end{figure*}

\begin{figure*}[t]
\centering{\includegraphics[width=132mm]{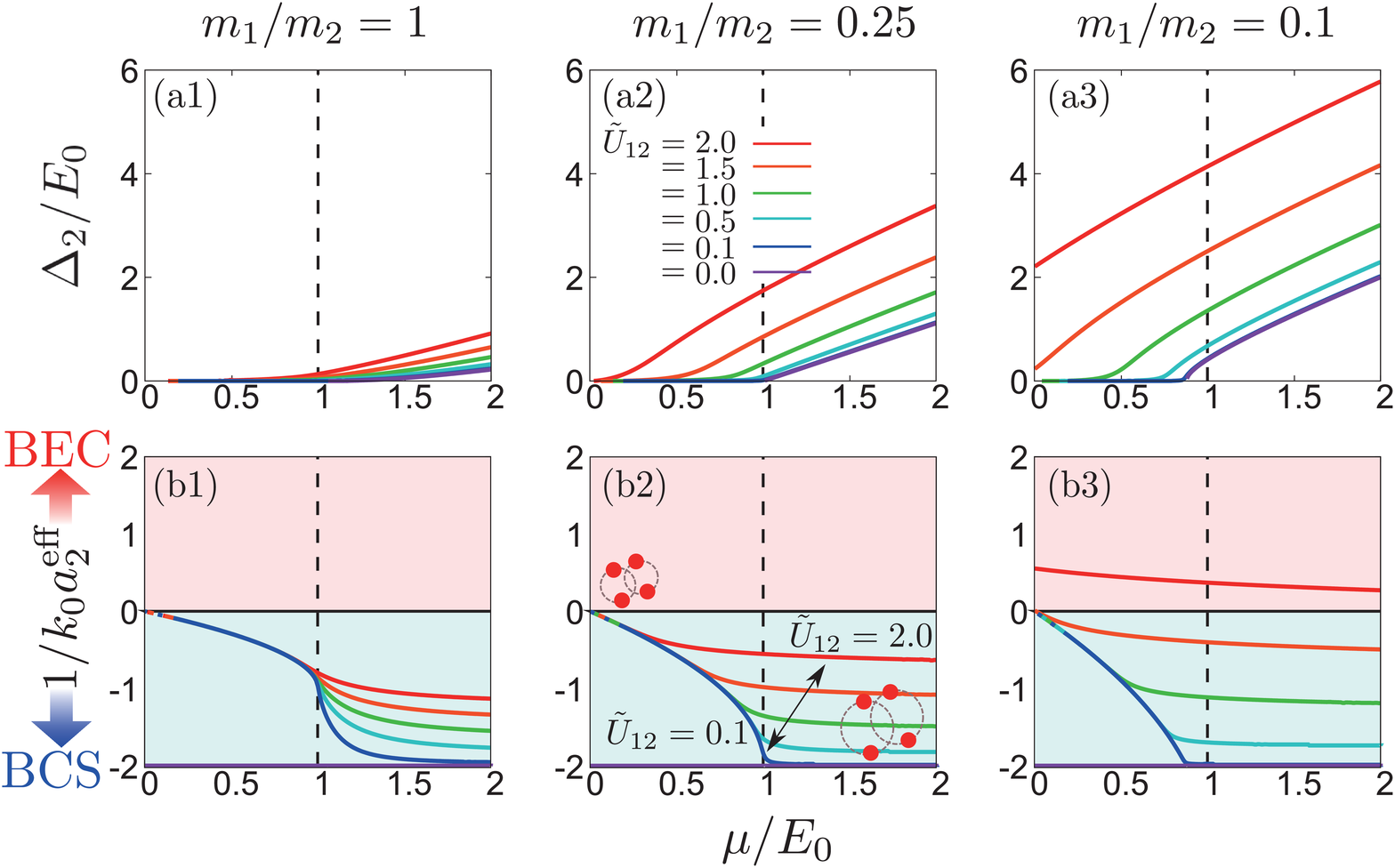}}
\caption{Same as Fig. \ref{fig3} for Band 2.
The behavior for $\mu \rightarrow 0$ with $1/(k_{0}a_{2}^{\rm eff})  \rightarrow 0$ is displayed as dotted lines following Eq. (\ref{eq:U22e}).}

\label{fig4}
\end{figure*}
\par
The result for the gap functions
against the chemical potential $\mu/E_0$,
calculated from the mean-field Eq.~(\ref{eq5}), is displayed in Fig.~\ref{fig3} for $\Delta_{1}$ and Fig.~\ref{fig4}
for $\Delta_{2}$ for the mass ratio $m_1/m_2 = 1, 0.25, 0.1$.
It is convenient to introduce a dimensionless
interband pair-exchange coupling~\cite{Yerin, Tajima},
\begin{eqnarray}
  \label{eq4}
\tilde{U}_{12} \equiv \left(\frac{\Lambda}{k_{0}}\right)^{2}\frac{n}{E_{0}}U_{12},
\end{eqnarray}
where $n=k_{0}^3/(3\pi^2)$ is
the total particle density as defined for
a zero-temperature ideal Fermi gas having a mass $m_1$
and a Fermi energy $E_{0}$.
For each value of $m_1/m_2$ we vary the interband
interaction $\tilde{U}_{12}$ from 0.0 to 2.0.
The result for the inverse effective scattering length \(1/a_{i}^{\mathrm{eff}}\), which serves as a
measure of the interaction strength, is also shown in the
lower panels of each figure.

\begin{figure*}[t]
\includegraphics[width=130mm]{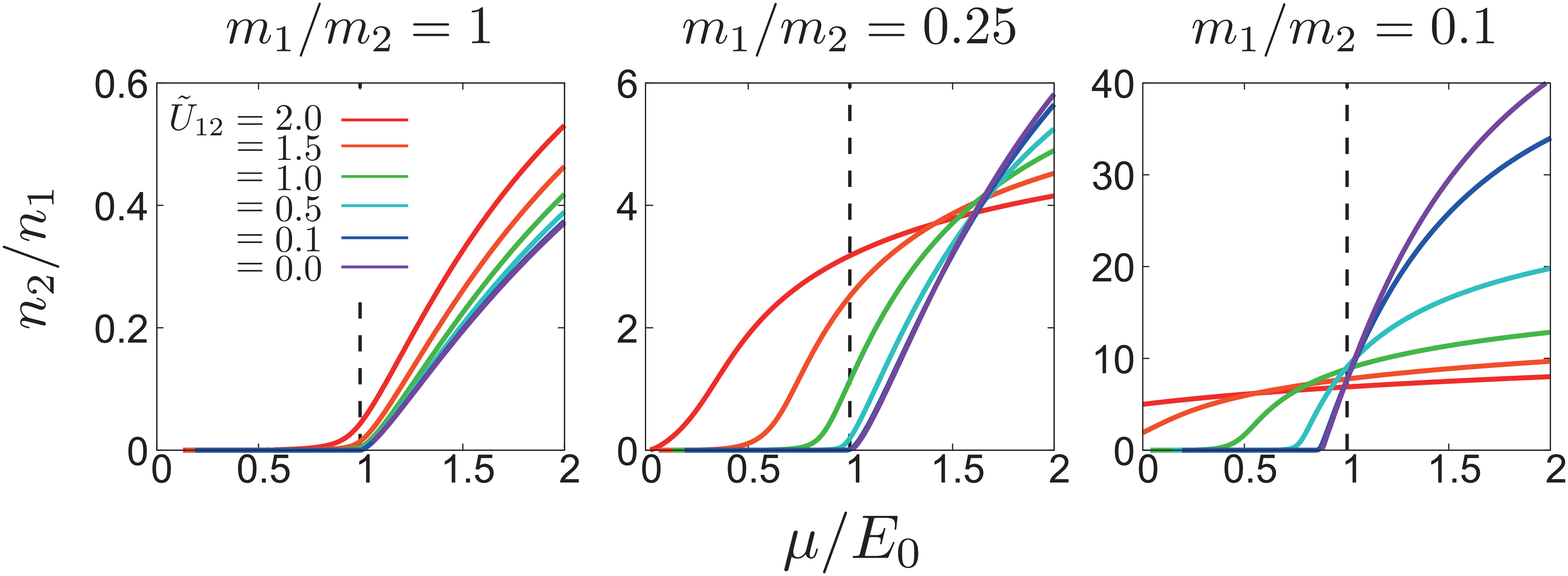}
\caption{Particle density ratio $n_2/n_1$  calculated for
the mass ratio \(m_{1}/m_{2}=1\) (left panel), \(m_{1}/m_{2}=0.25\) (middle), and \(m_{1}/m_{2}=0.1\) (right)
for various values of the pair exchange coupling $\tilde{U}_{12}$.  The vertical dashed lines mark $\mu=E_0$.
The intraband scattering lengths are set to $1/(k_{0}a_{1})=1/(k_{0}a_{2})=-2$ as in Figs.~3,~4.}
\label{fig5}
\end{figure*}

We can see that
both $\Delta_{1}$ and $\Delta_{2}$ increase with $\mu$,
but in a way vastly dependent on $m_1/m_2$ and $\tilde{U}_{12}$, both in their magnitude and the functional form
against $\mu$.
The enhancement of $\Delta_{1}, \Delta_{2}$ by the presence of the interband pair-exchange coupling $U_{12}$
can be regarded as a Suhl-Kondo mechanism~\cite{Suhl,Kondo}, but, crucially, this occurs more intensively
with orders of magnitude difference between $\Delta_{1}, \Delta_{2}$ for
larger mass difference (i.e., smaller mass ratio $m_1/m_2$),
as typically seen in the result for $m_1/m_2=0.1$ where the incipient band dispersion becomes almost flat.
\par

If we look at the band dependence more closely,
$\Delta_1$ is always nonzero, while
$\Delta_2$ vanishes for $\mu<E_{0}$ when $U_{12}=0$,
which is because Band 2 is unoccupied as depicted in
the result for the band occupancies in Fig.~\ref{fig5}
which shows that the number density ratio $n_2/n_1$
is virtually zero for $\mu<E_{0}$ in the absence of $U_{12}$ regardless of the value of $m_1/m_2$.
To be precise, even at $U_{12}=0$, the onset of nonzero density in Band 2 is slightly shifted toward the lower chemical potential with decreasing $m_1/m_2$,
a feature due to the intraband attraction.
\par

In the presence of $U_{12}$, on the other hand, $\Delta_{2}$ also becomes finite even for $\mu<E_{0}$.
There, $\Delta_{1}$ and $\Delta_{2}$ become simultaneously
finite through the coupling in Eq.~(\ref{eq5}) due to virtual pair-exchange processes.
Band 2 occupancy $n_2$ also becomes significantly
finite for $\mu<E_{0}$ due to $U_{12}$, implying the
acquisition of pair condensation in the incipient band
located above $\mu$.

Another characteristic feature is that both $\Delta_1$ and $\Delta_2$ remain finite even at $\mu=0$ when the mass ratio is small and the pair-exchange coupling is sufficiently large.  Although this may seem strange,
a bound state prevails in such a case as suggested in the context of a two-body problem.
In this regime, the pair formation originates from
the two-body bound state formation (as seen from
the pole of the \textit{T}-matrix discussed in Appendix ~\ref{ap1}) rather than the Cooper instability.
Indeed, we obtain finite two-body binding energies $E_{\rm bind}$ at $\tilde{U}_{12}=1.5$ and 2.0 therein.
The finite binding energy in the two-body problem is related to positive values of $1/(k_{0}a_{1}^{\rm eff})$ at $\mu=0$ in Fig.~\ref{fig3}~(b3).  However, we should note
that this argument does {\it not} hold for Band 2,
because $1/(k_{0}a_{2}^{\rm eff})$ for $\mu \rightarrow 0$
deviates from the result of Lippmann-Schwinger
equation due to the many-body effect as discussed in
Eq.~(\ref{eq:U22e}) below, i.e., $\Delta_{1}$ exerts
a significant effect
in Eq.~(\ref{eq92}) for $(i,j) = (2,1)$.
While $n_2$ increases with $m_2$ largely due to the
increased density of states,
the interband pair-exchange acts to reduce $n_2/n_1$ above $\mu=E_{0}$ as a result of the enhanced effective intraband attraction in Band 1, which we shall discuss below.
\par

\par

\subsection{Effective scattering length in each band\label{B}}
We have revealed in Figs.~\ref{fig3} and \ref{fig4} (b1-b3) that the inverse effective scattering lengths, $1/(k_{0}a_{1}^{\rm eff})$ and $1/(k_{0}a_{2}^{\rm eff})$ defined in Eq.~(\ref{eq8}),
have dramatically different dependence on the chemical potential
when we vary the mass ratio $m_1/m_2$.
Based on the result, we can now argue how the BCS-BEC crossover
evolves with $\mu$ in the present two-band model for various
values of the interactions $U_{ij}$.
The situation is indeed in a sharp contrast
with an ultracold Fermi gas around the magnetic Feshbach resonance where the BCS-BEC crossover can be realized by tuning the attractive interaction alone.
\begin{figure*}[t]
\centering{\includegraphics[width=180mm]{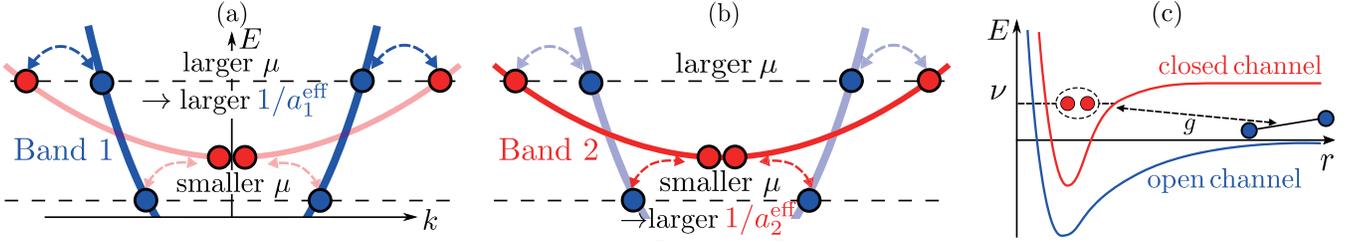}}
\caption{Conceptual correspondence between the present
two-band system (a, b) and the two-channel model(c) conventionally
used for the Feshbach resonance. In (a) for Band 1,
a virtual pair scattering from
around the Fermi energy in Band 1 to Band 2
are depicted in the momentum space. (b) depicts the pair scattering from
Band 2 to Band 1.
In (c)
the process of bound-state formation is depicted
for the closed and open channels against the relative coordinate ($r$) of two atoms, where $g$ is the Feshbach coupling, and $\nu$ is the energy level of the closed channel.
Then Band 2 can be analogous to a closed channel, although
Band 2 does not always form a bound state.}
\label{fig6}
\end{figure*}

\par
The effective intraband attraction in Band 1 as measured by \(1/(k_{0}a_{1}^{\mathrm{eff}})\) significantly and monotonically increases with
$\mu$, where the value changes from negative to positive
(i.e., $a_{1}^{\mathrm{eff}}$ itself {\it diverges})
typically in Fig.~3~(b2,~b3) for smaller mass ratios.
The sign change happens specifically around $\mu=E_{0}$ where $\mu$ touches the bottom of the incipient band.  We can capture its mechanism
as schematically depicted in Fig.~\ref{fig6}.
There, we compare the pair scattering processes in the
present two-band model with the
conventional magnetic Feshbach resonance in an ultracold single-atomic-species Fermi gas.  In the latter, the effective scattering length $a_{\rm eff}^{\rm Feshbach}$ between atoms in the two-channel model is given by~\cite{Ohashi}
\begin{eqnarray}
\label{eq:MFR1}
\frac{4\pi a_{\rm eff}^{\rm Feshbach}}{m_{\rm A}}
= U_{\rm bg}-\frac{g^2}{\nu}\frac{1}{1-(2\mu_{\rm A}/\nu)},
\end{eqnarray}
where $m_{\rm A}$ is the atomic mass, $\mu_{\rm A}$
the chemical potential, and $U_{\rm bg}$ the background interaction.
The attraction is induced by the Feshbach coupling $g$ between open-channel atoms and closed-channel molecules that have an energy level at $\nu$.
One can see in the above equation
that $a_{\rm eff}^{\rm Feshbach}$ diverges at $\mu_{A}=\nu/2$ due to the resonance tuned by $\mu_{A}$.
In the present two-band system,
$a_{1}^{\rm eff}$ diverges and changes sign
(with $a_2^{\rm eff}$ also rapidly changing;
see Fig.~\ref{fig4}) around $\mu=E_0$ for small $\tilde{U}_{12}$.
So we can regard this, where the
resonant pair scattering arises, as an analogue of
the Feshbach resonance accompanying a divergent $a_{\rm eff}^{\rm Feshbach}$.
To be more precise, the change of $1/a_{1}^{\rm eff}$ is related to the fact that $\Delta_{2}$ starts to increase around $\mu = E_{0}$ when $U_{12}$ is small (see Appendix \ref{ap2}).
Thus there exists an analogy between
the two-band system and the conventional Feshbach
resonance,
although there are some differences between the two models (such as the Feshbach resonance being described by the coupling between continuum and a bound state, whereas the resonant mechanism in the present two-band system originating from the coupling between two continua),
In this analogy, $U_{12}$ in the two-band model plays the role of $g$
in the magnetic Feshbach resonance.
So we can summarize the analogy as
\begin{table}[H]
\centering
\begin{tabular}{l|c|c}
 & present 2-band & Feshbach resonance \\
\hline
resonance energy & $\mu\simeq E_{0}$  &  $\mu_{A}=\nu/2$ \\
coupling        &  $U_{12}$ & $g$
\end{tabular}
\end{table}

\par
As exhibited conceptually in Fig.~\ref{fig6} (a) and numerically in Fig. \ref{fig3} (b3) for $\tilde{U}_{12}=2$, $1/(k_{0}a_{1}^{\rm eff})$ becomes large in a wide region of $\mu$ in contrast to the weak pair-exchange case when $U_{12}$ is large and $m_1/m_2$ is small.
Such a situation corresponds, in the present analogy, to the so-called
``broad Feshbach resonance" as
illustrated in Fig.~\ref{fig6}
in that $1/(k_{0}a_{\rm 1}^{\rm eff})$ is strongly enhanced as the interband interaction increases over a broad range of $\mu$ around $E_{0}$.
In this way, the Band 1 crosses from the weak-coupling BCS regime over
to the strong-coupling BEC regime with $\mu$ increasing across $E_{0}$ when $m_1/m_2$ is
sufficiently small and $\tilde{U}_{12}$ sufficiently large.
In particular, the effective interaction in Band 1
for large $U_{12}$ enters the strong-coupling regime even before $\mu$ reaches
the bottom of Band 2. Note that one of the important differences between the present two-band model and two-channel
atomic systems is the fact that Band 1 in the former
cannot be reduced to a single-channel model due to the large density of states in Band 2, which results in the enhancement of $n_2/n_1$ in Fig.~\ref{fig5} (a2, a3) for $\mu< E_{0}.$
For $\mu\gesim E_{0}$, on the other hand,
the strong effective interaction in Band 1 in that regime
acts to enhance $n_1$, hence reduce $n_2/n_1$ in Figs.~\ref{fig5} (a2, a3).

\par
The enhanced pairing effect associated with analogy between the two-band model and two-channel atomic system
occurs in both cases of the system coulpled with bosonic and fermionic bands. We also note that in Ref.~\cite{Avishai} a similar mechanism of the Feshbach resonance is proposed
for a two-body problem in a two-channel tight-binding model with equal effective masses.  There, a Feshbach resonance
in the long-wavelength limit
is discussed in terms of the scattering length and phase shift for varied one- and two-body potentials to reveal that the resonance can occur even when
the closed channel has no bound
states.  The present study, by contrast,
shows that the Feshbach analogue
arises driven by the chemical potential without changing any model parameters such as $U_{ij}$ and $E_0$.
Also, we study here a many-body system,
where a non-trivial realization of the unitarity limit in Band 2 in particular is induced by the coupled two superconducting order parameters, which would be outside a two-body scattering.
\par
If we turn to the incipient, heavy-mass Band 2, on the other hand,
the effective intraband interaction within the incipient band reaches the unitarity limit, that is $1/(k_{0}a_{2}^{\rm eff}) \rightarrow 0$ for $\mu \rightarrow 0$ in Fig.4.
We can also notice for the case of weak pair-exchange coupling
that the $\mu$-dependence of $1/(k_{0}a_{2}^{\rm eff})$
falls upon a universal behavior in the small $\mu$ limit
for various values of $\tilde{U}_{12}$.
This unitarity-limit behavior occurs as long as $\tilde{U}_{12}$ is nonzero
(note that $a_{2}^{\rm eff}=a_{2}$ for $\tilde{U}_{12}=0.0$).
In fact, we can
show in Appendix \ref{ap1} that,
whereas $1/(k_{0}a_{1}^{\rm eff})$ coincides with the two-body calculation at $\mu\rightarrow 0$ regardless of the value of $\tilde{U}_{12}$,
$1/(k_{0}a_{2}^{\rm eff})$ deviates significantly from the two-body calculation in the same limit in the presence of a nonzero $\tilde{U}_{12}$.
 This deviation stems from the coherent coupling between the binary condensates in the two-band system through the gap Eq.~(\ref{eq5}), from which we can rewrite $U_{22}^{\rm eff}$ as
\begin{eqnarray}
\label{eq:U22e}
U^{\rm eff}_{22}=\frac{1}{\sum_{\bm{k}}^{k\leq \Lambda}\frac{1}{2E_{2}(\bm{k})}}.
\end{eqnarray}
Note that the right-hand side of the above equation does not depend explicitly on $\tilde{U}_{12}$, a feature related to the aforementioned universal behavior of $1/(k_{0}a_{2}^{\rm eff})$ for small $\mu$ and $\tilde{U}_{12}$.
 At $\mu\rightarrow 0$ and $\Delta_2/E_{0}\simeq 0$, Eq.~(\ref{eq:U22e}) reduces to $U_{22}^{\rm eff}\simeq\left[\sum_{\bm{k}}^{k\leq \Lambda}\frac{1}{k^2/m_2+2E_{0}}\right]^{-1}$, leading to $1/a_{2}^{\rm eff}\rightarrow 0$ in Eq.~(\ref{eq8}).
This non-trivial realization of a unitarity limit in the incipient band can also be interpreted as a {\it narrow resonance} mechanism as opposed to the broad resonance,
where the ``narrow" means that the change of the effective scattering length occurs in a narrow range of the tuning parameter ($\mu$ in the present model,
a counterpart to $\nu$ in atomic systems); see more details in Appendix ~\ref{ap2}.
In other words, in the narrow resonance the
interband interaction ($g$) is weak, where the resonance
occurs abruptly in the vicinity of the resonce
condition.  Thus we can give a picture of
{\it the broad resonance for Band 1 with strong
interband interaction, and the narrow resonance
for Band 2 with weak interband interaction}.
 Incidentally, this situation does not apply when
the bound states are formed for small $m_1/m_2$ and
large $\tilde{U}_{12}$ as shown in Fig.~\ref{fig4}~(b3) ($\tilde{U}_{12}=2.0$, red line),
where the pairing is insensitive to the change of the chemical potential
as compared with the case of the Cooper instability
where the Fermi surface effect is crucial.
\par
On the other hand, when $\tilde{U}_{12}$ is small, the $1/(k_{0}a_{2}^{\rm eff})$ depends sensitively on the position of $\mu$ relative to $E_{0}$.
The qualitative behavior of \(1/(k_{0}a_{2}^{\rm{eff}})\)
around $\mu=E_{0}$ can again be
understood by analogy with the Feshbach resonance.
Namely, the light-mass band and the heavy-mass (incipient)
band correspond,
respectively, to the closed and open channels, as depicted in Fig.~\ref{fig6}~(b).
In the context of the atomic two-channel model,
assuming $\nu\rightarrow -|\nu|$,
[which corresponds to treating Band 2 as the
open channel in the two-channel model
described by Eq. (\ref{eq:MFR1})],
we obtain
\begin{eqnarray}
\label{eq:MFR2}
\frac{4\pi a_{\rm eff}^{\rm Feshbach}}{m_{\rm A}}=U_{\rm bg}+\frac{g^2}{|\nu|+2\mu_{\rm A}},
\end{eqnarray}
which indicates $4\pi a_{\rm eff}^{\rm Feshbach}/m_{\rm A}\rightarrow U_{\rm bg}$
for $\mu_{\rm A}\rightarrow\infty$.
 Correspondingly, by regarding the scattering continuum in Band 1 as the low-energy closed channel located at $-E_{0}$
below the Band 2 bottom,
and by identifying $U_{\rm bg}$ with the bare intraband interaction in our two-band model, we can again establish a
correspondence with the atomic model.
This way, one can obtain analogy in terms of
the effective scattering lengths between the two-channel model and the Feshbach resonance in atomic systems.
Indeed, despite various differences between the two models, $1/(k_{0}a_{2}^{\rm eff})$ still approaches $1/(k_{0}a_{2})$ (taken to be
$-2$ here)  for $\mu\gesim E_{0}$, as shown in Figs.~\ref{fig4} (b1-b3).
In this regard, the incipient band crosses from the unitarity limit
over to the weakly-coupling regime with increasing $\mu$,
which is just {\it opposite} to Band 1 where  $1/(k_{0}a_{1}^{\rm eff})$
increases with $\mu$.
When $\tilde{U}_{12}$ is large, Band 2 remains around the crossover even in the high-density regime ($\mu\gesim E_{0}$).

\par
\section{Conclusions \label{sec4}}
In this paper, we have investigated effects of resonant pair-exchange coupling and the resultant BCS-BEC and unitarity-BCS crossover in a two-band model consisting of dispersive and incipient nearly-flat bands.
Within the mean-field theory, we elucidate the chemical potential dependence of the superfluid/superconducting gaps and effective intraband interactions induced by the interband pair-exchange processes at various strengths of the pair-exchange coupling and effective mass ratio between the two bands.
We have found that superfluid/superconducting gaps in both bands are strongly enhanced when the incipient band becomes flat.
The effective scattering lengths which characterize the pair-exchange-induced effective attraction in the dispersive band are tuned from the weak-coupling to strong-coupling regimes only by increasing the chemical potential, leading to the BCS-BEC crossover without invoking any change in the coupling parameters.
We have discussed the analogy between the magnetic Feshbach resonance and the present two-band model in the presence of the incipient band.
Moreover, the nontrivial realization of the unitarity limit in the incipient band has been pointed out in the case of the small chemical potential, leading to
the unitarity-BCS crossover with increasing $\mu$.
\par
From an experimental point of view, while the effective scattering lengths cannot directly be measured in electronic systems,
the BCS-BEC crossover can be observed by measuring energy spectra in tunneling spectroscopies (STM/STS), which should exhibit quite different behaviors between the BCS and BEC regimes.  Moreover, ARPES (angular-resolved photoemission spectra) should give detailed information on quasiparticle spectra, as has actually been done for the iron-based superconductors for detecting a BCS-BEC crossover~\cite{Rinot}.

Although our model is rather simplified in describing real materials such as iron-based superconductors and bilayer graphenes,
our results would be useful for understanding strong-coupling properties of multi-band superfluid/superconductors.
Moreover, our approach could be applied to the topological flat band system as well as lattice models.


Thermal pairing fluctuations also play a crucial role throughout the BCS-BEC crossover.
These remain as important future work.


\acknowledgements
K. I. was supported by Grant-in-Aid for scientific Research from JSPS (Grant No. 18H01211). K. I. and H. T. were supported by Grant-in-Aid for scientific Research from JSPS (Grant No. 18H05406).
H.A. thanks Core Research for Evolutional
Science and Technology ``Topology" project from Japan Science
and Technology Agency, and JSPS KAKENHI (Grant JP17H06138).
\par
\appendix
\section{Comparison between two-body and many-body scattering properties~\label{ap1}}
{Here, we summarize two-body properties in the present two-band system.}
For convenience, we define a $2\times 2$ matrix $\hat{V}$
for the coupling constants in the band basis,
\begin{eqnarray}
  \label{eqa2}
  \hat{V} \doteq \left(
    \begin{array}{ccc}
      -U_{11} & -U_{12}  \\
      -U_{21} & -U_{22}
    \end{array}
  \right).
\end{eqnarray}
The in-vacuum two-body propagator is given by
\begin{eqnarray}
  \label{eqa3}
  \hat{J}(\omega_{+}) &\doteq& \left(
    \begin{array}{ccc}
      J_{1}(\omega_{+}) & 0  \\
      0 & J_{2}(\omega_{+})
    \end{array}
  \right),
\end{eqnarray}
where $\omega_{+}$ is the two-particle energy with an infinitesimal imaginary part +i$\delta$, and
\begin{eqnarray}
  J_{i}(\omega_{+})&=&\sum_{\bm{k}}^{k\leq\Lambda}\frac{1}{\omega_{+}-(k^{2}/m_{i}+2E_{0}\delta_{i2})}.
\end{eqnarray}

\begin{figure}[t]
\centering{\includegraphics[width=45mm]{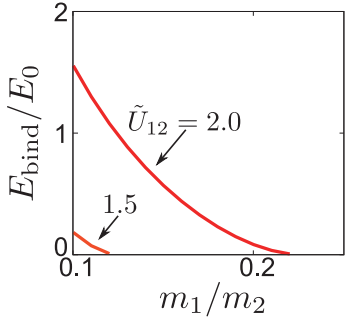}}
\caption{The binding energy $E_{\rm bind}$ calcurated
against $m_{1}/m_{2}$ for the $\tilde{U}_{12}=1.5$
and $2.0$.}
\label{fig7}
\end{figure}

We consider the diagonal component of the two-body 2$\times$2 $T$-matrix element $T_{i}(\omega_{+})$ in Band $i$, which is given by
\begin{eqnarray}
  \label{eqa4}
  T_{i}(\omega_{+})=\frac{\bar{U}^{\rm{eff}}_{ii}}{1-\bar{U}^{\rm{eff}}_{ii}J_{i}(\omega_{+})},
\end{eqnarray}
where
\begin{eqnarray}
  \label{eqa5}
  \bar{U}^{\rm{eff}}_{ii}=U_{ii}+U_{ij}\frac{J_{j}(\omega_{+})}{1-U_{jj}J_{j}(\omega_{+})}U_{ji}
\end{eqnarray}
is the two-body effective intraband interaction in Band $i$.
In the strong-coupling regime, we can obtain the two-body binding enegy $-E_{\rm bind}<0$ in Band 1 (which gives
$-E_{\rm bind}+2E_{0}$ for the two-body binding energy in
Band 2) from the pole of Eq.~(\ref{eqa4})
as
\begin{eqnarray}
\label{addeqa6}
1=\bar{U}_{ii}^{\rm eff}J_{i}(-E_{\rm bind}),
\end{eqnarray}
as shown in Fig.~\ref{fig7}.
The presence of a nonzero $E_{\rm bind}$ indicates that $\Delta_{1}, \Delta_{2}$ can be finite even at $\mu=0$ (as shown in Figs.~\ref{fig3}, \ref{fig4}).
\par
The low-energy limit $\omega_{+} \rightarrow 0$ of $\bar{U}^{\rm{eff}}_{11}$ coincides with Eq.~(\ref{eq9}) in the main text for $i=1$ at $\mu\simeq0$, since $\Delta_{2}/E_{0}\simeq 0$ even in the many-body counterpart.
On the other hand, $U_{22}^{\rm{eff}}$ does not coincide with $\bar{U}^{\rm{eff}}_{22}$.
More details about the deviation between $U_{22}^{\rm{eff}}$ and $\bar{U}^{\rm{eff}}_{22}$ are given in Appendix~\ref{ap2} below.
\par
We can further consider a situation in which the two-band system is in the BEC limit ($\mu<0$,~$\Delta_{1}, \Delta_{2}\ll |\mu|$) even when only the interband interaction exists with no intraband ones (large interband-coupling limit).
Equation (\ref{addeq10}) rewritten from Eq. (\ref{eq5}) is then approximated to
\begin{eqnarray}
\label{addeqa7}
1\simeq U_{ij}\sum_{\bm{k}}^{k\leq\Lambda}\frac{1}{k^{2}/m_{j}+2|\mu|+2E_{0}\delta_{2j}}U_{ji}\\ \nonumber
\times\sum_{\bm{k}}^{k\leq\Lambda}\frac{1}{k^{2}/m_{i}+2|\mu|+2E_{0}\delta_{i2}}
\end{eqnarray}
in the BEC limit for $i\neq j$.
The chemical potential in the BEC limit satisfies the same equation (\ref{addeqa6}) as that for the two-body binding energy in the absence of intraband interactions. Therefore, we obtain
\begin{eqnarray}
\label{addeqa8}
\mu=-\frac{E_{\rm bind}}{2}.
\end{eqnarray}
This equation is similar to the single-band case, where the chemial potential asymptotically approaches the result for half the two-body binding energy in the BEC limit at zero temperature.

\section{The low-density limit of \(\bm{1/(k_{0}a_{i}^{\mathrm{eff}})}\)}
\label{ap2}
\begin{figure*}[t]
\centering{\includegraphics[width=130mm]{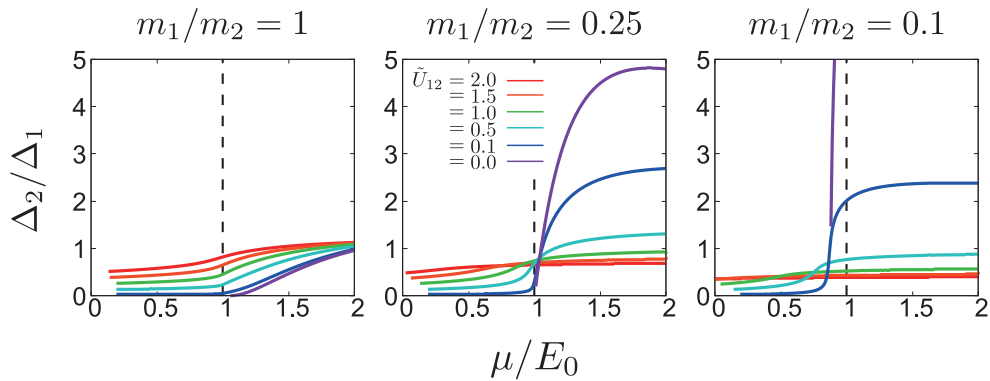}}
\caption{The ratio, \(\Delta_{2}/\Delta_{1}\), of
the gap functions as shown in Figs.~\ref{fig3} (a1-a3), Figs.~\ref{fig4} (a1-a3), is displayed for
the mass ratio \(m_{1}/m_{2}=1\) (left panel), \(m_{1}/m_{2}=0.25\) (middle), and \(m_{1}/m_{2}=0.1\) (right)
for various values of the pair exchange coupling $\tilde{U}_{12}$.  The vertical dashed lines mark $\mu=E_0$.}
\label{fig8}
\end{figure*}

Let us here clarify the mechanism by which
$1/(k_{0}a_{2}^{\rm{eff}})$ approach the unitarity limit despite the small pair-exchange interactions as long as $U_{12}$ is nonzero, while
$1/(k_{0}a_{1}^{\rm{eff}})$ is in the BCS regime for $\mu<E_{0}$.
\par
 First, note that $U_{ii}^{\rm{eff}}$ can be cast into
a form

\begin{eqnarray}
  \label{eqb1}
  U_{ii}^{\mathrm{eff}}
  =U_{ii}+U_{ij}\frac{\Delta_{j}}{\Delta_{i}}\frac{\sum_{\bm{k}}^{k\leq\Lambda}\frac{1}{2E_{j}(\bm{k})}}{\sum_{\bm{k}}^{k\leq\Lambda}\frac{1}{2E_{i}(\bm{k})}}
\end{eqnarray}
for $i \neq j$.
In the low-density region ($\mu<E_{0},~\Delta_{1}/E_{0}\simeq0,~\Delta_{2}/E_{0}\simeq0$),
Eq.~(\ref{eqb1}) becomes
\begin{eqnarray}
  \label{eqab2}
  U_{ii}^{\mathrm{eff}}&\simeq& U_{ii}+U_{ij}\frac{\Delta_{j}}{\Delta_{i}}\frac{\sum_{\bm{k}}\frac{1}{k^{2}/2m_{j}-\mu+E_{0}\delta_{j2}}}{\sum_{\bm{k}}\frac{1}{k^{2}/2m_{i}-\mu+E_{0}\delta_{i2})}}\\
  &=&U_{ii}+U_{ij}\frac{\Delta_{j}}{\Delta_{i}}
  \frac{m_{j}}{m_{i}}
  \frac{\tilde{\Lambda}+\frac{\sqrt{\frac{m_{j}}{m_{1}}(\tilde{\mu}-\delta_{j2})}}{2}\ln\left|\frac{\tilde{\Lambda}-\sqrt{\frac{m_{j}}{m_{1}}(\tilde{\mu}-\delta_{j2})}}
  {\tilde{\Lambda}+\sqrt{\frac{m_{j}}{m_{1}}(\tilde{\mu}-\delta_{j2})}}\right|}
  {\tilde{\Lambda}+\frac{\sqrt{\frac{m_{i}}{m_{1}}(\tilde{\mu}-\delta_{i2})}}{2}\ln\left|\frac{\tilde{\Lambda}-\sqrt{\frac{m_{i}}{m_{1}}(\tilde{\mu}-\delta_{i2})}}
  {\tilde{\Lambda}+\sqrt{\frac{m_{i}}{m_{1}}(\tilde{\mu}-\delta_{i2})}}\right|}
  ,\nonumber\\
  \label{eqab3}
\end{eqnarray}
where we have defined $\tilde{\mu} \equiv \mu/E_{0}$ and
$\tilde{\Lambda} \equiv \Lambda/k_{0}$.
Since we take a large cutoff such that $\sqrt{\tilde{\mu}}\ll\tilde{\Lambda},~\sqrt{\tilde{\mu}-1} \ll \tilde{\Lambda}$, we end up with
\begin{eqnarray}
\label{eqb4}
  U_{ii}^{\mathrm{eff}}&\simeq&U_{ii}+U_{ij}\frac{\Delta_{j}}{\Delta_{i}}\frac{m_{j}}{m_{i}}.
\end{eqnarray}
Hence $U_{22}^{\mathrm{eff}}$ depends strongly on the ratio $\Delta_{1}/\Delta_{2}$, while
$U_{11}^{\mathrm{eff}}$ depends conversely
on $\Delta_{2}/\Delta_{1}$.
As shown in Fig.~\ref{fig8}, $\Delta_{2}/\Delta_{1}$ for $\mu<E_{0}$ region becomes smaller as $\tilde{U}_{12}$ is
decreased at a given mass ratio.

\end{document}